\documentstyle[12pt]{article}
\renewcommand{\baselinestretch}{1.5}
\renewcommand{\arraystretch}{1.5}
\begin{document}
\thispagestyle{empty}
\pagestyle{empty}
\renewcommand{\thefootnote}{\fnsymbol{footnote}}
\newcommand{\preprint}[1]{\begin{flushright}
\setlength{\baselineskip}{3ex}#1\end{flushright}}
\renewcommand{\title}[1]{\begin{center}\LARGE
#1\end{center}\par}
\renewcommand{\title}[1]{\begin{center}\LARGE #1\end{center}\par}
\renewcommand{\author}[1]{\vspace{2ex}{\Large\begin{center}
\setlength{\baselineskip}{3ex}#1\par\end{center}}}
\renewcommand{\thanks}[1]{\footnote{#1}}
\renewcommand{\abstract}[1]{\vspace{2ex}\normalsize\begin{center}
\centerline{\bf Abstract}\par\vspace{2ex}\parbox{6in}{#1
\setlength{\baselineskip}{2.5ex}\par}
\end{center}}
\newcommand{\starttext}{\newpage\normalsize
\pagestyle{plain}
\setlength{\baselineskip}{4ex}\par
\setcounter{footnote}{0}
\renewcommand{\thefootnote}{\arabic{footnote}}
}
\newcommand{\segment}[2]{\put#1{\circle*{2}}}
\newcommand{\fig}[1]{figure~\ref{#1}}
\newcommand{\hc}{{\rm h.c.}}
\newcommand{\ds}{\displaystyle}
\newcommand{\eqr}[1]{(\ref{#1})}
\newcommand{\tr}{\,{\rm tr}}
\newcommand{\uone}{{U(1)}}
\newcommand{\su}[1]{{SU(#1)}}
\newcommand{\stu}{\su2\times\uone}
\newcommand{\be}{\begin{equation}}
\newcommand{\ee}{\end{equation}}
\newcommand{\bp}{\begin{picture}}
\newcommand{\ep}{\end{picture}}
\def\spur#1{\mathord{\not\mathrel{#1}}}
\def\lte{\mathrel{\displaystyle\mathop{\kern 0pt <}_{\raise .3ex
\hbox{$\sim$}}}}
\def\gte{\mathrel{\displaystyle\mathop{\kern 0pt >}_{\raise .3ex
\hbox{$\sim$}}}}
\newcommand{\sechead}[1]{\medskip{\bf #1}\par\bigskip}
\newcommand{\ba}[1]{\begin{array}{#1}\ds }
\newcommand{\cra}{\\ \ds}
\newcommand{\ea}{\end{array}}
\newcommand{\forto}[3]{\;{\rm for}\; #1 = #2 \;{\rm to}\; #3}
\newcommand{\for}{\;{\rm for}\;}
\newcommand{\cross }{\hbox{$\times$}}
\newcommand{\ol}{\overline}
\newcommand{\bra}[1]{\left\langle #1 \right|}
\newcommand{\ket}[1]{\left| #1 \right\rangle}
\newcommand{\braket}[2]{\left\langle #1 \left|#2\right\rangle\right.}
\newcommand{\braketr}[2]{\left.\left\langle #1 right|#2\right\rangle}
\newcommand{\g}[1]{\gamma_{#1}}
\newcommand{\half}{{1\over 2}}
\newcommand{\del}{\partial}
\newcommand{\grad}{\vec\del}
\newcommand{\real}{{\rm Re\,}}
\newcommand{\imag}{{\rm Im\,}}
\newcommand{\gapprox}{\raisebox{-.2ex}{$\stackrel{\textstyle>}
{\raisebox{-.6ex}[0ex][0ex]{$\sim$}}$}}
\newcommand{\lapprox}{\raisebox{-.2ex}{$\stackrel{\textstyle<}
{\raisebox{-.6ex}[0ex][0ex]{$\sim$}}$}}
\newcommand{\cl}[1]{\begin{center} #1\end{center}}
\newcommand\etal{{\it et al.}}
\newcommand{\prl}[3]{Phys. Rev. Letters {\bf #1} (#2) #3}
\newcommand{\prd}[3]{Phys. Rev. {\bf D#1} (#2) #3}
\newcommand{\npb}[3]{Nucl. Phys. {\bf B#1} (#2) #3}
\newcommand{\plb}[3]{Phys. Lett. {\bf #1B} (#2) #3}
\newcommand{\ie}{{\it i.e.}}
\newcommand{\etc}{{\it etc.\/}}
\def\cA{{\cal A}}
\def\cB{{\cal B}}
\def\cC{{\cal C}}
\def\cD{{\cal D}}
\def\cE{{\cal E}}
\def\cF{{\cal F}}
\def\cG{{\cal G}}
\def\cH{{\cal H}}
\def\cI{{\cal I}}
\def\cJ{{\cal J}}
\def\cK{{\cal K}}
\def\cL{{\cal L}}
\def\cM{{\cal M}}
\def\cN{{\cal N}}
\def\cO{{\cal O}}
\def\cP{{\cal P}}
\def\cQ{{\cal Q}}
\def\cR{{\cal R}}
\def\cS{{\cal S}}
\def\cT{{\cal T}}
\def\cU{{\cal U}}
\def\cV{{\cal V}}
\def\cW{{\cal W}}
\def\cX{{\cal X}}
\def\cY{{\cal Y}}
\def\cZ{{\cal Z}}
\renewcommand{\baselinestretch}{1.5}
\renewcommand{\arraystretch}{1.5}
\newcommand{\boxit}[1]{\ba{|c|}\hline #1 \\ \hline\ea}
\newcommand{\mini}[1]{\begin{minipage}[t]{20em}{#1}\vspace{.5em}
\end{minipage}}

\preprint{\#HUTP-92/A036\\ 7/92}
\title{
Generalized Dimensional Analysis\thanks{Research
supported in part by the National Science Foundation under Grant
\#PHY-8714654.}\thanks{Research supported in part by the Texas National
Research Laboratory Commission, under Grant \#RGFY9206.}
}
\author{
Howard Georgi \\
Lyman Laboratory of Physics \\
Harvard University \\
Cambridge, MA 02138
}
\date{}
\abstract{
I describe a version of so-called naive dimensional analysis, a rule for
estimating the sizes of terms in an effective theory below the scale of chiral
symmetry breaking induced by a strong gauge interaction. The rule is
simpler and more general than the original, which it includes as a special
case. I also give a simple qualitative interpretation of the rule.
}

\starttext

In dealing with effective field theories describing physics of mesons below a
symmetry breaking scale in a strongly interacting theory, it is important to
have a tool for estimating the coefficients of nonrenormalizable interactions.
 Naive dimensional analysis (NDA)~\cite{nda} was proposed as such a tool. It
works pretty well in QCD. However, a better instrument is needed for theories
in which the number of colors and flavors may be very different from what they
are in QCD. In this very brief note, I describe one. I will give a rule for
such dimensional estimates that is both simpler and more general than the
original. This simplicity and generality is obtained by introducing an
additional parameter, the ratio of the Goldstone boson decay constant to the
mass of the lightest non-Goldstone bound-states. I will also give an extremely
simple, qualitative argument to interpret the rule.

I will consider only strongly interacting theories that are ``QCD-like'' ---
with fermions transforming only under the simplest representation of the gauge
group, in order to avoid the additional complications of chiral fermions and
of dependence on ratios
of Casimir operators. For example, I don't want to think about
``tumbling''~\cite{tumbling} because it makes my head hurt.~\cite{headache}
The low energy physics described by these QCD-like theories is the physics of
the light pseudo-Goldstone mesons. The effective field theory is only useful
at energies small compared to the scale at which other bound-states appear.

At the end, I speculate on the dependence of the extra parameter on
the color and flavor structure of the theory and discuss possible
generalizations of the trivial idea described here. The discussion in this
paper is very simple. I do not pretend that it is very deep. It has probably
been stated in only slightly different form by others. Nevertheless, I think
that the very simplicity of the statement is a virtue. It strips NDA down to
its barest essentials. I think that this is useful in trying to determine the
form the dimensional analysis will take in more interesting effective
theories.

In order to be able to keep track of things like the number of colors and
flavors in the various strong groups, I will distinguish the Goldstone boson
decay constant, $f$, from the typical mass of the low-lying (non-Goldstone)
bound states, $\Lambda$. In QCD, $f=f_\pi$ is the Goldstone boson decay
constant
and $\Lambda\approx 1$GeV (or the $\rho$ mass --- take your pick) is the
typical mass of the light but non-Goldstone bound states.

The simple rule to assign a dimensional coefficient of the right size to any
term in the effective Lagrangian is
\be\parbox{.8\hsize}{
\begin{enumerate}
\item
include an overall factor of $f^2\Lambda^2$;
\item
include a factor of $1/f$ for each strongly interacting field;
\item
add factors of $\Lambda$ to get the dimension to 4.
\end{enumerate}}
\label{rule}\ee
It is that simple. The mass $\Lambda$ now implicitly contains the factor of
$4\pi$ from the original version of NDA. In fact, this simple rule encompasses
all the cases discussed in the original version of NDA, including external
fields, quark masses, and the like.

However, this rule also makes it possible to extend NDA to different
numbers of colors, for example. If $N_c$ is large in QCD, then as $N_c$
changes, $f$ scales with $N_c^{1/2}$ while $\Lambda$ does not change. The
result of \eqr{rule} then agrees with more sophisticated analyses so long as
you are calculating the coefficient of a term that is leading in powers of
$N_c$.\footnote{You can, of course, foul up any scheme for estimating sizes by
looking at a term that is suppressed by some symmetry.}~\cite{improvednda}

Even for theories in which the large $N_c$ arguments do not apply (QCD may be
in this class as well --- we don't know for sure \cite{reconsidered}), this
formulation of dimensional analysis makes sense.
The cost of this increased generality is that we now have an additional
parameter, $\Lambda/f$, to fix before we can use dimensional analysis. Except
for QCD, where we can read the answer from the particle data book, we do not
really know $\Lambda/f$.

Why should it work? The idea is simple. What can the coefficients depend on?
They will clearly depend on the masses of the non-Goldstone bound-states.
Nonrenormalizable interactions among the Goldstone bosons can be produced by
virtual exchange of these bound-states, so their momentum dependence, at
least, will presumably be set by $\Lambda$. If this were the whole story,
there would be no more story. However, we know that it is not. We need an
independent parameter, $1/f$, that measures the amplitude for making a
Goldstone boson. The rule, \eqr{rule}, is just the statement that these two
effects are the only things going on. There are no other large or small
parameters in the strongly interacting theory. There is a dimensional scale
set by the strong interactions --- this is $\Lambda$. There is an amplitude
for emitting a Goldstone boson --- either the dimensional constant, $1/f$, or
if you prefer, a dimensionless number, $\Lambda/f$, one for each Goldstone
boson, and the rest is simply dimensional analysis with the mass scale
$\Lambda$.

I should say that there is absolutely nothing special about the Goldstone
bosons in this analysis except that they are light. If we are willing to
extend the effective theory to describe other meson states as well, we
would expect coefficients consistent with exactly the same rule. Similarly,
the same rule applied to baryons gives the conventional NDA result that each
dimension 3/2 baryon field gets a factor of ${1\over f\sqrt{\Lambda}}$.
Thus I interpret the inverse Goldstone boson decay constant as a more or less
universal measure of the amplitude for producing a strongly interacting bound
state. Fields describing weakly interacting particles, on the other hand,
behave just like external fields and are suppressed by the appropriate factors
of $1/\Lambda$. Of course, unless there is some reason for the other strongly
interacting states to be light, it is not obvious that effective theories
describing these heavier states would be of much use.~\cite{ineffective}

We do not know very much for certain about the dependence of the ratio,
$\Lambda/f$ on the number of flavors, $N_f$, and colors, $N_c$, in the gauge
theory. The bound from the original NDA,
\be
{\Lambda\over f}\lte4\pi \,,
\label{general}\ee
must still be satisfied, because the arguments of \cite{nda} and
\cite{weinberg} are still valid, but in general there will be stronger
constraints. For example,
we know that for sufficiently large $N_f$ and $N_c$, it goes like
$1/N_c^{1/2}$ times some function of the ratio, $N_f/N_c$.~\cite{veneziano}
The authors of \cite{reconsidered}, elaborating an argument of Kaplan, suggest
that the ratio has the form:
\be
{\Lambda\over f}\approx\min\left(
{4\pi a\over N_c^{1/2}},
{4\pi b\over N_f^{1/2}}\right)\,,
\label{form}
\ee
where $a$ and $b$ are constants of order 1. This is a useful provisional form.
Note that \eqr{general} is satisfied.

If this generalization of NDA proves to be useful, we will want to know how to
generalize it further. Can anything similarly simple be said about chiral
gauge theories in which the low energy effective theory contains light
fermions as well as bosons? There are many uncertainties in this kind of
generalization. How do the scales depend on the Casimir operators? Is the
analog of $f$ for the production of chiral fermions the same as for bosons?
These questions are interesting and difficult field theory. If nature chooses
to make use of chiral gauge theories above the $\su2\times\uone$ breaking
scale, they may one day become relevant phenomenology.

\section*{Acknowledgements} I am grateful to Sekhar Chivukula, David Kaplan,
and Aneesh Manohar for interesting comments.

\end{document}